\documentclass[conference]{IEEEtran}
\IEEEoverridecommandlockouts
\usepackage{cite}
\usepackage{amsmath,amssymb,amsfonts}
\usepackage{algorithmic}
\usepackage{graphicx}
\usepackage{textcomp}
\usepackage{xcolor}
\usepackage{braket}
\usepackage{listings}
\def\BibTeX{{\rm B\kern-.05em{\sc i\kern-.025em b}\kern-.08em
    T\kern-.1667em\lower.7ex\hbox{E}\kern-.125emX}}
\begin{document}

\title{Cache Blocking Technique to Large Scale Quantum Computing Simulation on Supercomputers}

\author{\IEEEauthorblockN{1\textsuperscript{st} Jun Doi}
\IEEEauthorblockA{\textit{IBM Quantum, IBM Research - Tokyo} \\
\textit{}\\
Tokyo, Japan \\
doichan@jp.ibm.com}
\and
\IEEEauthorblockN{2\textsuperscript{nd} Hiroshi Horii}
\IEEEauthorblockA{\textit{IBM Quantum, IBM Research Tokyo} \\
\textit{}\\
Tokyo Japan \\
horii@jp.ibm.com}
}

\maketitle

\begin{abstract}
Classical computers require large memory resources and computational power to simulate quantum circuits with a large number of qubits. 
Even supercomputers that can store huge amounts of data face a scalability issue in regard to parallel quantum computing simulations because of the latency of data movements between distributed memory spaces. 
Here, we apply a cache blocking technique by inserting swap gates in quantum circuits to decrease data movements. 
We implemented this technique in the open source simulation framework Qiskit Aer. 
We evaluated our simulator on GPU clusters and observed good scalability.
\end{abstract}


\section{Introduction}
Hundreds of qubits of quantum computers are becoming realistic in the near future and existing quantum computers are available for development and evaluation of quantum applications \cite{aspuru2005simulated,kandala2017hardware,o2016scalable}. 
Some of these quantum computers are available as cloud services, but their computer resources are still not enough or their accesses are limited. 
As well, numbers of qubits will still be limited and noise of quantum gate operations will exist in this few decades. 
Therefore, quantum computing simulations running on classical computers are necessary for developing and debugging new quantum applications as well as actual quantum hardware is. 
However, quantum computing simulators on classical computers require huge amounts of memory and computational resources. 
An $n$-qubit state vector requires an array of length $2^n$ to be stored on a classical computer; for example, a $50$-qubit simulation needs 16 petabytes worth of storage for double precision floating point numbers. 
This means that simulating large numbers of qubits on a classical computer is a form of supercomputing.

In this study, we focus on accelerating quantum computing simulations by using GPUs on hybrid distributed parallel computers. 
Universal quantum computing simulations \cite{qiskit,q-sharp,project-q} are now available for developing quantum applications with smaller numbers of qubits (around $20$ qubits) on classical computers, even on desktop or laptop personal computers. 
To simulate rather more qubits (around $50$-qubits), parallel simulators \cite{li2017quantum,haner20170,jones2018quest,smelyanskiy2016qhipster,doi2019} must store the quantum state in the huge distributed memory in parallel-processing computers. 
Parallel computing has the advantage of accelerating simulations that require a huge amount of computational power. 
Graphic processor units (GPUs) are particularly useful in this aspect: their thousands of hardware threads update probability amplitudes in parallel and their high bandwidth memory (HBM) reduces bottlenecks in loading and storing the amplitudes.
Some quantum computing simulators \cite{qulacs,tyson2019,doi2019,qiskit-aer} support the use of GPUs to accelerate simulations. 

We focus on optimizing a state vector simulator, which is simple and stable to simulate noisy and noiseless quantum computing simulations on classical parallel computers.
There are some approaches to decrease the memory usage of a quantum computing simulation on classical computers. 
For example, data compression specialized for quantum computing simulation were introduced in \cite{DeRaedt2019,wu2019} to simulate noisy quantum computers with more than $50$ qubits on supercomputers. 
However, their lossy data compression and lacks of fidelity limit quantum algorithms to be simulated. 
Use of tensor network \cite{Villalonga_2020} is another approach to simulate noiseless quantum computers while
storing quantum state as tensor matrices.
Tensor-network-based simulators can reduce memory for specific applications where qubits are weekly interacted but not for general applications where qubits are strongly interacted.
We believe that a state vector simulator will be the most popular in developing and debugging quantum algorithms even though it requires huge memory to directly store $2^n$ probability amplitudes.
Users can simply increase more available qubits with additional computing nodes and data storage.

Though system can provide enough memory, inefficient memory usage in state vector simulators is a critical problem to simulate many qubits; the communication overheads over hybrid parallel computers become inhibitors to scale simulation performance. 
To parallelize a state vector simulator on a distributed memory space, probability amplitudes of quantum state must be exchanged across different memory spaces. 
In general, network bandwidth between different memory spaces is much lower than memory bandwidth of CPU and GPU.
Minimizing data exchanges across memory spaces is a key to scale simulation performance.

Here, we propose a new technique to decrease the number of data exchanges across distributed memory spaces by combining quantum circuit optimization and parallel optimization. 
\textit{Qubit reordering or remapping} are circuit optimizations used to map circuits onto the topology of quantum devices.
\textit{Qubit reordering} can be used to decrease data exchanges between large numbers of qubit gates. 
The technique we developed decreases data exchanges by moving all the gates associated with a smaller number of qubits by inserting noiseless swap gates. 
The operations of the optimized circuits resembles cache blocking on a classical computer, and the concept of the optimization is similar to one on a classical computer because we block the gates on the qubits that can be accessed faster. 

Section \ref{sec:parallel_sv} overviews quantum computing and state vector simulators. 
In Section \ref{sec:parallel_sv}, we describe how to parallelize state vector simulators on distributed computers, and in Section \ref{sec:cache_block}, we describe the circuit optimization and cache blocking implementation of a parallel state vector simulator. 
In Section \ref{sec:eval}, we discuss performance evaluations on a hybrid parallel computer accelerated by GPUs. We conclude the paper in Section \ref{sec:summary} with mention of future work.

\section{Parallel State Vector Simulation}
\label{sec:parallel_sv}

\subsection{State Vector Simulation Overview}
We focus on simulating universal quantum computers based on quantum circuits that consist of one-qubit rotation gates and two-qubit CNOT gates. These gates are known to be universal; i.e., any quantum circuit (for realizing some quantum algorithm) can be constructed from CNOT and one-qubit rotation gates.

Quantum circuits run quantum computing programs by applying quantum gates to qubits. A qubit has two basis states, $\ket{0}=\begin{pmatrix}1\\0\end{pmatrix}$ and $\ket{1}=\begin{pmatrix}0\\1\end{pmatrix}$. While a classical bit stores $0$ or $1$ exclusively, a qubit allows superposition of the two basis states as $\ket{\psi}=a_0\ket{0}+a_1\ket{1}$, where the probability amplitudes $a_0$ and $a_1$ are complex numbers satisfying $|a_0|^2+|a_1|^2=1$. When a qubit is measured, an outcome of $0$ or $1$ is obtained with probability $|a_0|^2$ or $|a_1|^2$, respectively. This means two complex numbers for tracking the quantum state of a one-qubit register must be stored in the simulation.

The quantum state of an $n$-qubit register is a linear superposition of $2^n$ basis states, each of the form of the tensor product of $n$ qubits. For example, the basis state of digit `2' (or $10$ in binary) can be represented as $\ket{2}=\ket{1}\otimes\ket{0}=\ket{10}=(0,0,1,0)^T$, where $T$ denotes the transpose of a vector. Thus, the quantum state of the $n$-qubit register can be written as $\ket{\psi}=\sum_{i=0}^{2^n-1}a_i\ket{i}$. Note that each state $\ket{i}$ has its own probability amplitude $a_i$ in a complex number. Overall, simulating quantum circuits requires that these $2^n$ complex numbers be stored to enable tracking of the evolution of the quantum state of an $n$-qubit register. We denote these complex numbers as `$\mathsf{qreg}$.'

Quantum gates transform the quantum state of the $n$-qubit register by rotating its complex vector. We focus on the gate set defined by the OpenQASM specification \cite{cross2017open}. The set consists of an arbitrary single-qubit (rotation) gate (a u3 gate) and a two-qubit gate (controlled-not or CNOT or CX). The u3 gates are rotations of the 1-qubit state and are mathematically defined as

\begin{equation}
\label{eq:u3}
 u3(\theta, \psi, \lambda) =
 \begin{pmatrix}
 	\cos(\theta/2)&-e^{i\lambda}\sin(\theta/2)\\
 	e^{i\psi}\sin(\theta/2)&e^{i(\psi+\lambda)}\cos(\theta/2)
 \end{pmatrix}.
\end{equation}

When it is applied to the $k$-th qubit of $\{q_n \cdots q_1\}$, a u3 gate transforms the probability amplitudes of the basis states $\newline\ket{x_n \cdots x_{k+1}\ 0\ x_{k-1}\cdots x_1}$ and $\ket{x_n \cdots x_{k+1}\ 1\ x_{k-1}\cdots x_1}$ for each $x_i$ in $\{0, 1\}$ by linear transformation of their previous probability amplitudes. This implies that applying a single-qubit u3 gate can change all $2^n$ complex numbers stored for tracking the evolution of the quantum state. However, the changes can be computed in parallel, as each of the new probability amplitudes depends on its own value and on the value of a probability amplitude whose basis state is different at the $k$-th location.

The CNOT gates are applied to two qubits: a control and a target qubit. If the control qubit is in state $\ket{1}$, the CNOT gate flips the target qubit. If the control qubit is in state $\ket{0}$, the CNOT gate does not change the target qubit. If we have two qubits and take the higher bit as the control qubit and the other as the target qubit, the CNOT gate is mathematically defined as
\begin{equation}
\label{eq:cnot}
 \mbox{CNOT} =
 \begin{pmatrix}
 1&0&0&0\\
 0&1&0&0\\
 0&0&0&1\\
 0&0&1&0\\
 \end{pmatrix}.
\end{equation}
When applied to the $c$-th and $t$-th qubits (control and target, respectively), the CNOT gate swaps the probability amplitude of $\ket{x_n \cdots (x_c=1) \cdots (x_t=0) \cdots x_1 }$ with that of the $\ket{x_n \cdots (x_c=1) \cdots (x_t=1) \cdots x_1 }$. The swaps, which affect half of the probability amplitudes of the quantum state, can also be performed in parallel.

\begin{figure}[tbp]
  \centering
  \includegraphics[width=4cm]{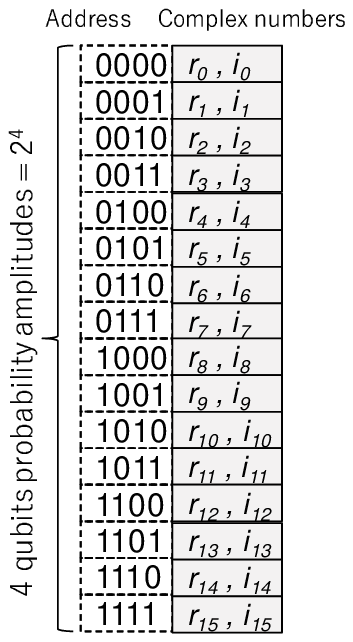}
\caption{Example of 4-qubits state vector array on a classical computer. The array has $2^4 = 16$ complex numbers to store probability amplitudes.}
\label{fig:qb_array}
\end{figure}

On a classical computer, the state vector is stored as an array of probability amplitudes which are expressed as floating point complex numbers. Fig. \ref{fig:qb_array} shows an example of a state vector array of a 4-qubit quantum circuit stored in the memory of a classical computer. If we store a complex number as a double precision floating point number, an $n$-qubit state vector requires $16\times2^n$ bytes of memory on a classical computer. 

\begin{figure}[tbp]
  \centering
  \includegraphics[width=\linewidth]{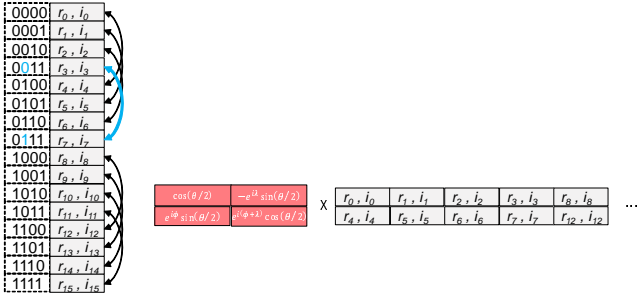}
\caption{$u3$ gate applied to $2$-nd qubit (zero-based index). Pairs of probability amplitudes whose distance is $2^2$ are selected to be rotated by multiplying it with a $2 \times 2$ matrix. }
\label{fig:u3_gate}
\end{figure}

Quantum gate operations can be simulated on a classical computer by multiplying a matrix with pairs of probability amplitudes stored in the state vector array. Fig. \ref{fig:u3_gate} shows an example of a $u3$ gate, which rotates the $2$-nd qubit (zero-based index) on a 4-qubit state vector stored in the memory of a classical computer. A $2 \times 2$ matrix is multiplied by pairs of probability amplitudes, which are selected by applying the $XOR$ operation to their addresses; e.g. binary indices $0011$ and $0111$ become a pair through $0011 \oplus 2^{k=2} = 0111$. The $u3$ gate constructs pairs from all of the probability amplitudes in the state vector, meaning that all the elements in the state vector are accessed and updated to simulate a gate operation. This requires $byte/flop \fallingdotseq 2.29$ in double precision, so simulating a gate operation on a classical computer is a memory-bandwidth bound problem.

\subsection{Chunk-based Parallelization}
In parallel computing environment, distributed processes in nodes and GPUs have own memory spaces.
Only when data is allocated to a memory space, its process can perform calculation on it.
To maximize parallelization, data should not be frequently exchanged across spaces in general.
On the other hand, saving memory is very important to store $2^n$ probability amplitudes.
Once their size exceeds capacity of a memory space, some amplitudes are spilled out to slow disk storage. 

\begin{figure}[tbp]
  \centering
  \includegraphics[width=\linewidth]{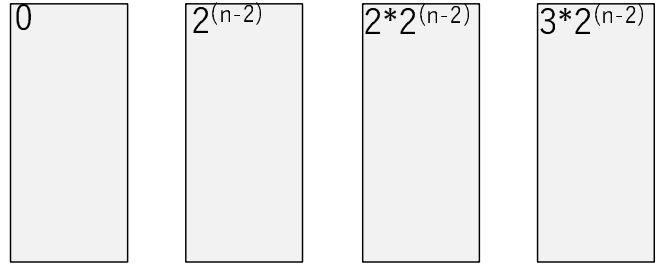}
\caption{Distributing an $n$-qubit state vector to 4 distributed memory spaces. The number in each box shows the beginning index of the probability amplitude in each sub-array. }
\label{fig:sv_dist}
\end{figure}

We divide a state vector array into sub-arrays $chunk$s and allocate them to distributed memory space without duplication.
Fig. \ref{fig:sv_dist} shows an example where an $n$-qubit state vector is divided to four distributed memory spaces with four chunks of $2^{n-2}$ probability amplitudes.
Gate operations on qubits from $0$ to $n-3$ are performed within a process and data exchange is necessary in operations for $n-2$ and $n-1$ qubits.

\begin{figure}[tbp]
  \centering
  \includegraphics[width=\linewidth]{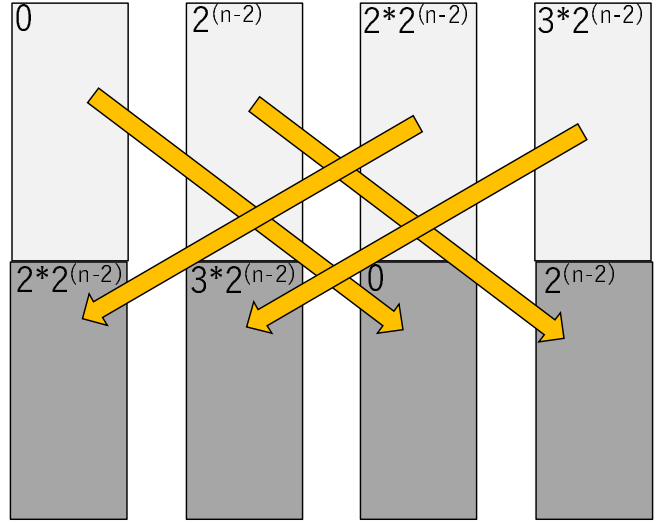}
\caption{Naive implementation of probability amplitude exchange between distributed memory spaces. In this example, we apply a gate to qubit $k=n-1$. This implementation requires twice the memory space}
\label{fig:naive_exchange}
\end{figure}

Fig. \ref{fig:naive_exchange} is a naive way to do this; it has a buffer to receive a copy of the probability amplitudes from remote memory. 
This implementation is very simple; however, it requires double the memory space to have a copy of the remote memory and hence is not memory efficient. 
This hinders us from increasing the number of qubits to be simulated.

\begin{figure}[tbp]
  \centering
  \includegraphics[width=\linewidth]{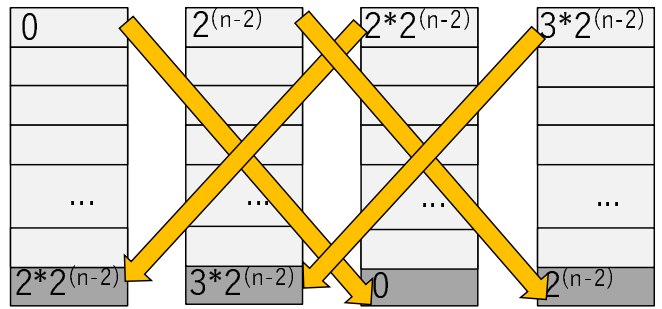}
\caption{Dividing a state vector into small chunks and performing probability amplitude exchange by chunks. So we only need additional one chunk per memory space to perform probability amplitude exchange }
\label{fig:chunk_exchange}
\end{figure}

We use a chunk as a unit of data exchange. 
Double memory spaces are necessary for buffering if we divide a state vector with large chunks as shown in Fig \ref{fig:naive_exchange}.
Therefore we divide the state vector into $2^{nc}$ small chunks and exchange them in a pipeline to minimize space for buffering.
Fig. \ref{fig:chunk_exchange} shows an example to use only one buffer to exchange chunks.

In summary, a gate operation of a chunk on $m$ memory spaces has three types based on a target qubit $k$: if $k$ is lower than $nc$, it can be independently processed, and if $k$ is lower than $ \log_n m$, referring other is necessary, otherwise, the chunk is exchanged. 
By minimizing gate operations on qubits of more than equal to $ \log_n m$, data exchange across memory spaces are reduced.

\section{Implementation of State Vector Simulator}
\label{sec:sv_impl}

\subsection{Qiskit Aer Overview}
Qiskit Aer \cite{qiskit-aer} is an open source framework for quantum circuit simulations and we implemented a parallel quantum computing simulator for high-performance computers as a backend of Qiskit Aer.
Qiskit Aer is one of the components of Qiskit \cite{qiskit} that provides seamless access to quantum hardware and simulators with libraries to help development of quantum software. 
In addition to ideal (noiseless) simulation, Qiskit Aer supports a noise model for noisy simulation, which allows users to develop quantum algorithms for realistic noisy environments.
The frontend of Qiskit-Aer translates such Qiskit-unique APIs to basic matrix operation as written in Section \ref{sec:parallel_sv} for backends.
Depending on characteristics and usages of quantum circuits, users select one of backends from state vector, unitary matrix, density matrix, stabilizer state, super operator, and matrix product state (MPS) simulators.
Out backend extends the state vector simulator with acceleration of GPUs and distributed environment.

\subsection{Accelerating Simulation by GPUs}
Qiskit Aer provides a state vector simulator accelerated by using GPUs, which is the basis of our backend for GPUs and MPI.
Since Qiskit Aer is written in C++14 and the standard template library (STL), its GPU codes use the Thrust library \cite{Thrust}, which is a template class library optimized for GPU acceleration. 
Thrust provides readability and productivity in addition to performance; its templates are more readable than CUDA \cite{cuda} and compilable to GPU, OpenMP, and Intel's Threading Building Blocks (TBB) \cite{TBB} binaries for various host computers.  

The Thrust library is based on the vector class it resembles \verb|std::vector| class of STL.
The library includes GPU-ready optimized operations on vectors such as copying, transforming, sorting, and reduction.
The API supports simple lamda equations that work in CPU and GPU environment.
The List \ref{lst:multiply–add} is an example of a customized kernel in Thrust that calculates $y = a\times x + y$ in double precision.
Inside the \verb|thrust::transform|, runtime of Thrust automatically load and store elements of two vectors \verb|x_dev| and \verb|y_dev| while evaluating a defined binary operator \verb|daxpy(a)| to update corresponding elements of \verb|y_dev|.

\begin{figure}
\begin{lstlisting}[caption=An example of multiple-accumulation written in Thrust, basicstyle=\ttfamily\footnotesize, frame=single, captionpos=b, label=lst:multiply–add]
struct daxpy {
  double a;
  daxpy(double a_in) {a = a_in;}
  //user defined binary operation
  __host__ __device__ 
  double operator()(double& x,double& y)
  {return a*x+y;}
};

thrust::host_vector<double> x(n), y(n);
thrust::device_vector<double> x_dev(n), y_dev(n);
//copy vector from host to GPU
x_dev = x;
y_dev = y;

thrust::transform(thrust::device, 
  x_dev.begin(),x_dev.end(),y_dev.begin(),daxpy(a));

//copy back result from GPU to host
y = y_dev;
\end{lstlisting}
\end{figure}

Implementing quantum gate operations become more complex than List \ref{lst:multiply–add}.
Unlike \verb|thrust::transform| with a binary operator \verb|daxpy(a)|, two elements of a vector are accessed and updated non-sequentially.
Thus, to load and store a pair of probability amplitudes, we use \verb|thrust::for_each| and a unary operator.
\verb|thrust::for_each| just calls the operator with an index number in a specified range and the unary operator does not return any. 
List \ref{lst:u3} shows an implementation of a u3 gate by using Thrust.
In \verb|operator()| function, two elements in a vector \verb|pVal| that store probability amplitudes are identified as \verb|pVal[i0]| and \verb|pVal[i1]| and  updated by multiplying a unitary matrix (\verb|m00|, \verb|m01|, \verb|m10| and \verb|m11|).
We iterates this function with the half size of the vector.

\begin{figure}
\begin{lstlisting}[caption=A kernel of u3 gate written in Thrust, basicstyle=\ttfamily\footnotesize, frame=single, captionpos=b, label=lst:u3]
struct u3_gate {
  thrust::complex<double>* pVec;
  thrust::complex<double> m00,m01,m10,m11;
  int add;
  int mask;
  void u3_gate(thrust::complex<double>* pV,
        thrust::complex<double>* pM,int k){
    pVec = pV;
    m00 = pM[0]; m10 = pM[1];
    m01 = pM[2]; m11 = pM[3];
    add = 1ull << k;
    mask = add - 1;
  }
  __host__ __device__ 
  void operator()(const uint_t &i) const {
    int i0,i1;
    thrust::complex<double> q0,q1;
    //calculate address from index
    i1 = i & mask;
    i0 = (i - i1) << 1;
    i0 += i1;
    i1 = i0 + add;
    q0 = pVec[i0];
    q1 = pVec[i1];
    pVec[i0] = m00 * q0 + m01 * q1;
    pVec[i1] = m10 * q0 + m11 * q1;
  }
};

thrust::device_vector<double> chunk(size)
thrust::complex<double>* pV;
auto ci = thrust::counting_iterator<int>(0);
pV = thrust::raw_pointer_cast(chunk.data());
thrust::for_each(thrust::device,ci,ci+size/2,
                  u3_gate(pV,mat,k));
\end{lstlisting}
\end{figure}

\section{Cache Blocking Technique to Parallel State Vector Simulator}
\label{sec:cache_block}

\subsection{Transpiling Circuit for Cache Blocking}
Since performance of the state vector simulator is bounded by memory-bandwidth, reduction of data movements is essential to shorten simulation time.
In hybrid parallel computing, two types of memory spaces exist; CPU and GPU.
Therefore, we list five types of data movements to be minimized for simulator performance.

\begin{itemize}
\item Loading data from the memory of a CPU
\item Copying data from the memory of a CPU to the memory of a GPU
\item Loading data from the memory of a GPU
\item Copying data from one GPU to another GPU
\item Copying data to other nodes
\end{itemize}

Cache blocking is a well-known technique to avoid repeating fetch data from main memory to CPU caches in high-performance computing research. 
We extends this technique to avoid frequent data movements of the above by reusing data stored in local fast memory as long as possible. 
If consecutive gates perform qubits that are smaller than the number of chunk bits $nc$, any data movements do not occur while simulating them.
We transform (transpile) a circuit to have many of long sequences of such consecutive gates for simulator performance.

If some qubits are not updated in any gates, we can use bit reordering, which is a transpilation technique to map gates onto the actual quantum computing hardware. 
We map such qubits to qubits larger than or equal to $nc$ and other qubits to smaller qubits than $nc$ in simulation.
Though this technique can reduce data movements, we cannot remove all the data movements if a circuit updates $nc$ or larger qubits.

\begin{figure}[tbp]
  \centering
  \includegraphics[width=\linewidth]{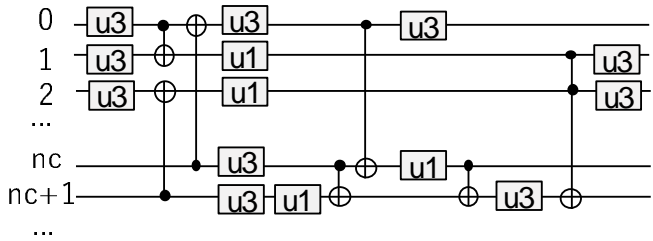}
\caption{Example of an input quantum circuit consisting of u1, u3 and CNOT gates. $nc$ denotes the number of qubits of a chunk. The gates on qubits > nc should refer probability amplitudes over multiple chunks to be simulated. }
\label{fig:input_circuit}
\end{figure}

\begin{figure}[tbp]
  \centering
  \includegraphics[width=\linewidth]{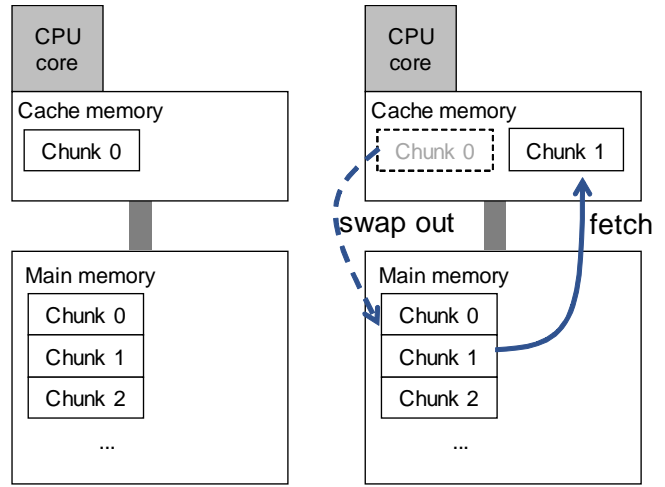}
\caption{Memory hierarchy on a classical computer. Cache memory is placed near the CPU core; it is very fast, but its size is limited. In this example, only one chunk can be stored in cache. So if we fetch a chunk from main memory, the older chunk has to be swapped out.}
\label{fig:classic_cache}
\end{figure}

\begin{figure}[tbp]
  \centering
  \includegraphics[width=\linewidth]{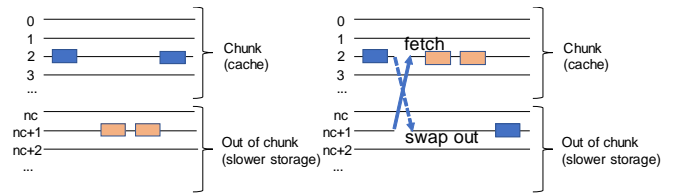}
\caption{Cache memory of quantum circuit on the state vector simulator. Gates on the smaller qubits (smaller than $nc$) can be calculated without moving data, so the memory is very fast, like a cache memory on a classical computer. A swap gate can be used to fetch and swap out a pair of qubits. }
\label{fig:quantum_cache}
\end{figure}

To reduce more data movements, we use another hardware mapping technique that inserts swap gates to move some gates to other qubits. 
Fig. \ref{fig:input_circuit} shows an example of a circuit in which all the qubits are updated. We assume that gates on qubits larger than or equals to $nc$ require data movements between chunks. Similar to the cache memory on a classical computer, a chunk whose size is $nc$ qubits on a quantum computer can be cached as shown in Fig. \ref{fig:classic_cache}. The qubits in excess of $nc$ can be assigned to slower memory, and by applying a swap gate to a pair of qubits, one qubit is fetched to the cache and the other qubit is swapped out from cache at the same time (Fig. \ref{fig:quantum_cache}). Then, gate operations on the fetched qubit can be performed in the cache.  

\begin{figure}[tbp]
  \centering
  \includegraphics[width=\linewidth]{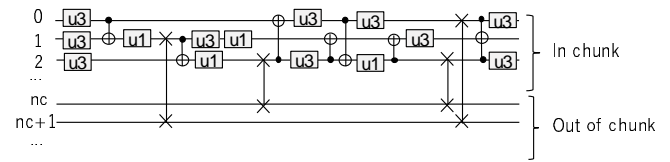}
\caption{Example of a cache blocked circuit. Four swap gates are added to move all the gates to fewer qubits than $nc$, now all the gate can be performed without referring probability amplitudes over chunks.}
\label{fig:swapped_circuit}
\end{figure}

Before simulating a given quantum circuit, we apply transpiler to move all the gates inside a chunk by inserting swap gates. Fig. \ref{fig:swapped_circuit} shows the circuit transpiled from the input circuit shown in Fig. \ref{fig:input_circuit}.

The circuit consists of a sequence of gates. First, we collect the gates that can be executed in a chunk by reordering them. The execution order of two gates can be swapped if each gate is applied to different qubits because these two gates can be executed independently on a real quantum computer. However, a pair of gates in a multi-qubit gate such as CNOT can not be reordered if the pair of gates are laid on the same qubits. As well, all the qubits used in a multi-qubit gate should be placed in a chunk at the same time. Thus, multi-qubit gates are used to choose which qubits are to be stored in a chunk. 

List \ref{lst:cacheblock} describes the algorithm of the cache blocking transpiler. First, it searched for the multi-qubit gates from the sequence of gates and chooses $nc-1$ qubits that are used in these multi-qubit gates. Then, noiseless swap gates are inserted to put the chosen qubits in a chunk. These are special swap gates, named \verb|chunk_swap|, that are different from the usual swap gate. Then, the gates are reordered by putting those which can be executed in a chunk. If there is a multi-qubit gate which is laid across a chunk, this gate prevents the remaining gates on the same qubits from being executed. So we set these qubits as blocked qubits, and gates with blocked qubits are stored in queue. We repeat this procedure until the remaining gates are not empty.

\begin{figure}
\begin{lstlisting}[caption=traspilation for cache blocking, basicstyle=\ttfamily\footnotesize, frame=single, captionpos=b, label=lst:cacheblock]
put sequence of gates in queue REMAINING
while REMAINING is not empty
  choose QB_CHUNK[nc-1] qubits to store in chunk
  put noiseless swap_chunk gates in queue OUTPUT

  clear QB_BLOCKED[nc-1]
  for gates in REMAINING
    if gate's qubits are in QB_CHUNK 
             and not QB_BLOCKED is set
      put gate in queue OUTPUT
    else
      if multi-qubit gate 
             and some qubits are in QB_CHUNK
        for gate qubits
          set QB_CHUNK
      put gate in queue REMAINING_NEXT
  copy REMAINING_NEXT to REMAINING
\end{lstlisting}
\end{figure}

We also insert special gates to specify a section of gates to be blocked in a chunk. These \verb|begin_blocking| and \verb|end_blocking| gates are inserted to specify a section, and the gates in the section can be simulated independently in each chunk. 

The algorithm we are proposing in this section is kind of an heuristic algorithm and it is not the best optimized solution because we are applying the cache blocking transpilation on the fly before the actual simulation so it is not good to take long time to get the best optimized transpilation. If we can transpile the circuit outside of the simulation, there can be more optimal algorithms to decrease chunk swaps. 

\subsection{Data Structure of Parallel Qiskit Aer}
\begin{figure}[tbp]
  \centering
  \includegraphics[width=\linewidth]{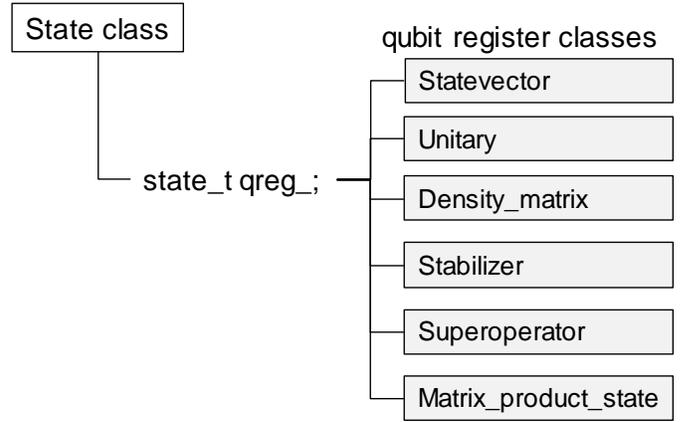}
\caption{Data structure of state class in serial Qiskit Aer. A state has one qubit register class that has a quantum state with the corresponding simulation method.}
\label{fig:aer_state}
\end{figure}

As described in section \ref{sec:sv_impl}, Qiskit Aer has simulation methods, and these methods are defined in the classes with a quantum register which stores the quantum state. One execution of a quantum circuit, called a \verb|"shot"|, is controlled in the \verb|State| class that has one qubit register class for the corresponding simulation method, as shown in Fig. \ref{fig:aer_state}. 

\begin{figure}[tbp]
  \centering
  \includegraphics[width=\linewidth]{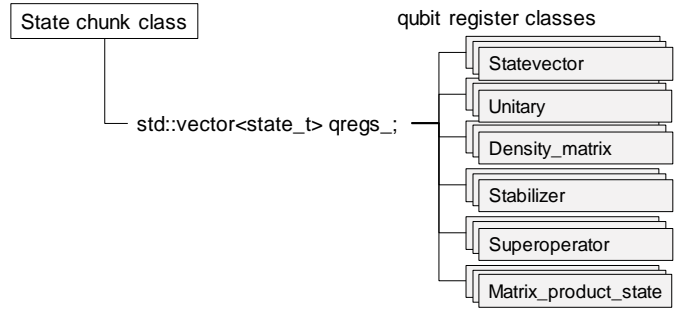}
\caption{Extension of state class to multiple chunks for parallel Qiskit Aer. A chunk is defined on the basis of the usual qubit register classes. }
\label{fig:aer_state_chunk}
\end{figure}

To parallelize a simulation, we need to have multiple chunks in a state class; hence, we extend the \verb|State| class to the \verb|State_chunk| class that has multiple chunks in an array, as shown in Fig. \ref{fig:aer_state_chunk}. Each chunk can be implemented on the basis of existing qubit register classes, with small changes, because we do not have to communicate between chunks when simulating gates other than the swap gates inserted by the cache blocking transplier. Each qubit vector class does not have to know where they are in the whole $n$-qubit state; they only have to be simulated as if the state has $nc$ qubits.

\subsection{Parallel Implementation}
The \verb|State_chunk| class manages multiple chunks and controls the simulation by handling a sequence of gates. There are $2^{n-nc}$ chunks in total to simulate an $n$-qubit circuit, and if we have $p$ processes, we distribute the chunks to all of these processes. We add continuous chunk indices to each chunk and can get the beginning address of the probability amplitude for each chunk from $chunk index\times2^{nc}$. To initialize a state to $\ket{0}$, we only initialize the chunk with index $0$ to $\ket{0}$ and set all other chunks to zero. 

\begin{figure}[tbp]
  \centering
  \includegraphics[width=\linewidth]{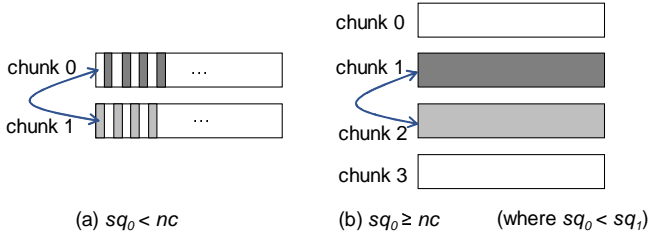}
\caption{Two cases of chunk\_swap applications: (a) half of the amplitudes in the chunks are swapped; (b) two chunks of four combined chunks are swapped. The latter case only occurs when we reorder qubits for the output. }
\label{fig:chunk_swap}
\end{figure}

List \ref{lst:chunkupdates} describes how we simulate with multiple chunks in the \verb|State_chunk| class. We scan three special gates, \verb|begin_blocking|, \verb|end_blocking| and \verb|chunk_swap| inserted by the cache blocking transplier. The sequence of gates placed between \verb|begin_blocking| and \verb|end_blocking| can be independently simulated in each chunk, so we can parallelize the loop of the chunks by mapping to multiple GPUs or multiple threads on a CPU.

\begin{figure}
\begin{lstlisting}[caption=update multiple chunks, basicstyle=\ttfamily\footnotesize, frame=single, captionpos=b, label=lst:chunkupdates]
for gate in sequence of gates
  if gate is begin_blocking
    for all chunks
      fetch chunk on GPU if stored on CPU
      for gate in blocking section
        apply gate
      swap out chunk to CPU
  else if gate is chunk_swap
    for all pairs of chunks
      apply swap
  else
    for all chunks
      apply gate
\end{lstlisting}
\end{figure}

We use the memory of the GPUs as a cache to simulate a sequence of gates in the blocking section when some of the chunks are stored on the memory of a CPU. Since the performance of gate simulation is bounded by the memory bandwidth, the high memory bandwidth of GPUs is an advantage. However, the bandwidth between the CPU and the GPUs is much narrower than the memory bandwidth of the CPU memory, so the performance is much degraded from that of a CPU if a chunk is transferred to simulate only one gate on a GPU. By using this cache blocking technique, we can take the advantage of the GPU's high memory bandwidth if we have enough gates in the blocking section. 

A \verb|chunk_swap| gate is applied to a pair of chunks. As in the case of the usual swap gate, the \verb|chunk_swap| gate takes two qubits $sq_0$ and $sq_1$ (where $sq_0<sq_1$) to be swapped, and one or two of the target qubits are larger or equal to chunk qubit $nc$. Fig. \ref{fig:chunk_swap} shows two cases of chunk pairing: (a) is when $sq_0<nc$; here, a pair of chunks is selected and half of the amplitudes are swapped between chunks. (b) is when $sq_0\geq nc$; here, four chunks are selected and all the amplitudes in two chunks \verb|chunk 1| and \verb|chunk 2| are swapped. In both cases, we swap amplitudes between two chunks, so we select two chunks by referring to $sq_0$ and $sq_1$ and apply \verb|chunk_swap|. Also, if the selected chunks are on different processes, we have to exchange the chunks by using MPI communication. We send a chunk to the other process that has the other chunk and receive a chunk from the same process. We replace the amplitudes in the chunk with those of the received chunk but we do not return the chunk to the pairing process.

We added the \verb|chunk_swap| function to the qubit register classes that takes a reference to the pairing qubit register class as a parameter to swap amplitudes between the chunks selected in \verb|State_chunk| class. The implementation of this function depends on the simulator; the paragraph below describes the implementation of the \verb|chunk_swap| function for the state vector simulator. 

\begin{figure}[tbp]
  \centering
  \includegraphics[width=\linewidth]{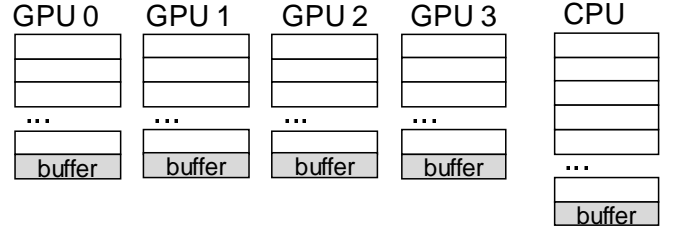}
\caption{Chunk allocation in a process with multiple GPUs. If we can not store all the chunks in the memory of one GPU, we distribute the chunks to multiple GPUs. If some of the chunks can not be stored in the GPUs, we store them in the memory of the CPU.}
\label{fig:chunks_1node}
\end{figure}

One chunk whose size is $2^{nc}$ is stored in the qubit register class for the state vector simulator. If there are GPUs installed on the computer, we try to store the chunks in the memory of one of the GPUs. If the chunks cannot be stored in the memory of one GPU, we try to store them in two or more GPUs. If the memories of all the GPUs become full, we store the remaining chunks in the main memory of the CPU, as shown in Fig. \ref{fig:chunks_1node}. To swap chunks stored in different memory spaces, i.e. chunks on different GPUs or chunks on a GPU and chunks on a CPU, we have copy chunks into the buffer to swap amplitudes in the same memory space. As a result, we have to allocate buffer chunks before storing chunks in memory. 

Since the raw data of a chunk depends on the simulation method, the qubit register classes prepare chunk data to be sent to the pairing process by using MPI\_Send. A buffer to receive a chunk from the pairing process by using MPI\_Recv is also prepared by the qubit register classes. For the state vector simulator, a chunk itself is used to send to the pairing process and a buffer allocated on the memory space is used for a receive buffer.

\section{Performance Evaluation}
\label{sec:eval}

\subsection{Computer Environment}
We compared the simulation times of a state vector simulator implementing the cache blocking technique with that of the simulator not implementing the technique on the GPU cluster (See Table \ref{tab:env_for_eval} \cite{Vetter18} \cite{tesla-v100}). Each node of the cluster had six GPUs, and each GPU had 16GB of memory, so we could simulate up to 29 qubits on a single GPU with double precision. The size of the main memory was 512GB, on which we could simulate up to 34 qubits. Moreover, by storing chunks in the memories of both the GPUs and CPU as shown in Fig. \ref{fig:chunks_1node}, we could simulate up to 35 qubits on a single node.

\begin{table}[tbp] 
\caption{Computer environments used in the evaluation of the state vector simulator} 
\label{tab:env_for_eval}
\hbox to\hsize{\hfil
\begin{tabular}{l|l}\hline\hline
    Cluster node & IBM Power System AC922 \\
    CPU & POWER9 \\
    Number of sockets per node & 2 \\
    Number of cores per socket & 21 \\
    CPU memory size & 512GB \\
    GPU & NVIDIA Tesla V100 \\
    Number of GPUs per node & 6 \\
    GPU memory size & 16GB \\
    CPU - GPU interconnect & NVLink2 \\
    Inter connect & Infiniband EDR\\
    OS & Red Hat Enterprise Linux Server 7.6 \\
    Compiler & GCC 8.3.0 \\
    CUDA Toolkit & CUDA 10.1 \\
    MPI & IBM Spectrum MPI 10.3.1\\\hline
\end{tabular}\hfil}
\end{table}

The base implementation without the cache blocking technique is described in \cite{doi2019} and chunks are exchanged when the target qubits of the gates are larger or equals to chunk qubit. The base implementation was implemented as the set of a \verb|State| and qubit register classes shown in Fig. \ref{fig:aer_state}, so parallelization was managed inside the qubit vector class. We implemented the kernels of gates by using Thrust on both implementations. 

For the evaluation we prepared two quantum circuits. One of the circuit we used is Quantum Volume \cite{Andrew2019} which is a randomly generated quantum circuit that uses all the given qubits and all the qubits entangled by CNOT gates. The Quantum Volume circuit itself does not have a mathematical meaning but is useful for evaluating quantum systems. The other circuit is Quantum Fourier Transform (QFT) which is one of the most important quantum algorithms and occurs in many quantum applications. As in Quantum Volume, there are entanglements over all the qubits but the other calculations are multiplication of diagonal unitary matrices. On the state vector simulator diagonal matrix multiplications can be calculated independently on each probability amplitudes, so data exchange between chunks of QFT is much less than that of Quantum Volume. 

\subsection{Single Node Performance Comparison}
\begin{figure}[tbp]
  \centering
  \includegraphics[width=\linewidth]{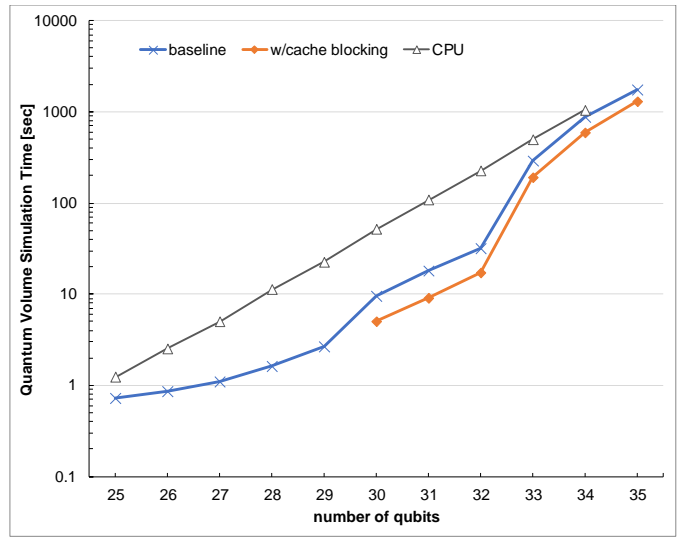}
\caption{Single node simulation time of Quantum Volume (depth=10) measured on IBM Power System AC922 with 6 GPUs. Only one GPU is used $ \leq 29 $ qubits, and we can store all data in memory of six GPUs $\leq 32$ qubits. Both memory of GPU and CPU are used $ \geq 33 $ qubits. }
\label{fig:singlenode}
\end{figure}

\begin{figure}[tbp]
  \centering
  \includegraphics[width=\linewidth]{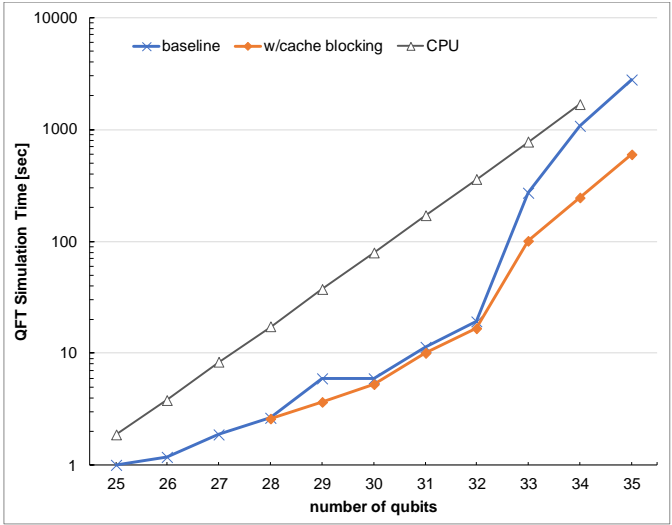}
\caption{Single node simulation time of QFT measured on IBM Power System AC922 with 6 GPUs. Only one GPU is used $ \leq 29 $ qubits, and we can store all data in memory of six GPUs $\leq 32$ qubits. Both memory of GPU and CPU are used $ \geq 33 $ qubits.}
\label{fig:singlenode_qft}
\end{figure}

Fig. \ref{fig:singlenode} and Fig. \ref{fig:singlenode_qft} show the measured simulation time of Quantum Volume and QFT circuits on a single node of an IBM Power System AC922 (Table \ref{tab:env_for_eval}) equipped with six NVIDIA Tesla V100 GPUs. They also show the time on a CPU for comparison. While the simulation time on the CPU is linear, we can see steps in the simulation times of the GPU implementations (baseline and w/cache blocking). 

The first step from 25 to 29-qubits shows the simulation time calculated by one GPU; there is no data exchange between GPUs, so the simulation time with cache blocking is not plotted because it is the similar to the baseline. On 28 and 29-qubits in Fig. \ref{fig:singlenode_qft}, we also plotted the simulation time by using multiple GPUs with cache blocking. We got better performance than using one GPU even if we have to exchange chunks between GPUs because there are fewer data exchange occurs, it seems the calculation time shortened by parallelization by multiple GPUs overcomes the overheads of data exchange. 

The second step from 30 to 32 qubits shows the simulation times when using six GPUs; here, the performance deteriorates on the baseline of Quantum Volume because of data exchange overheads, on the other hand with cache blocking we can decrease overheads of data exchange and we measured better performance. The performance of QFT on the second step shows different behavior. Because there is less data exchange occurs on QFT, the performance of the baseline did not deteriorate, so there is small advantage in cache blocking here.

The third step from 33 to 35 qubits shows the simulation times when chunks are stored in the memories of the six GPUs and the CPU. Since the CPU has 512GB of memory, we can simulate up to 34 qubits on it, but by storing chunks on both the GPUs and CPU, we can simulate 35 qubits. Thus, the cache blocking technique reduces the data exchange overhead, offering a speedup of about 20\% from the baseline for Quantum Volume and 80\% for QFT. In the third step, the baseline implementation calculates the chunks on the CPU when the chunks are stored in the memory of the CPU; on the other hand, the cache blocking implementation transfers chunks to the GPU to calculate the sequence of gates. The GPU cache blocking also leads to a performance gain.

\subsection{Scalability Comparison}
\begin{figure}[tbp]
  \centering
  \includegraphics[width=\linewidth]{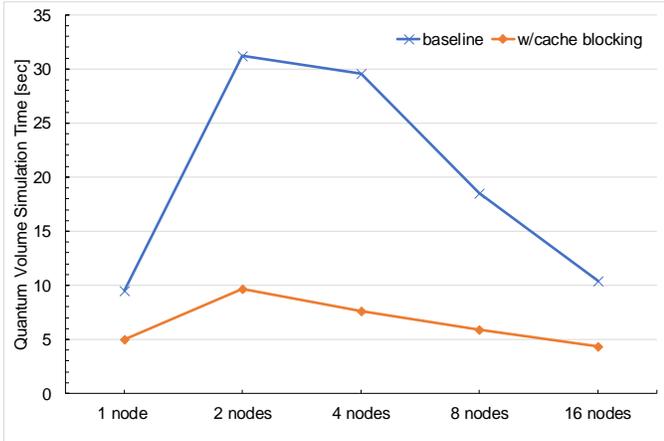}
\caption{Simulation time of 30-qubit Quantum Volume (depth=10) on IBM Power System AC922 with 6 GPUs (strong scaling). For 1 node, we used all 6 GPUs but we only used 1 GPU per node for the larger number of node cases.}
\label{fig:strong_30qubits}
\end{figure}

\begin{figure}[tbp]
  \centering
  \includegraphics[width=\linewidth]{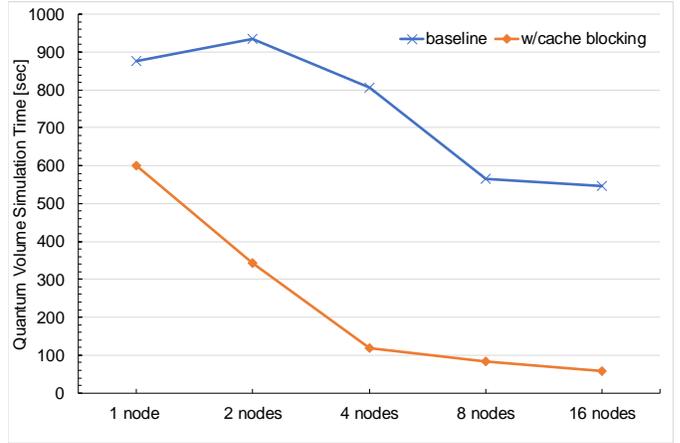}
\caption{Simulation time of 34-qubit Quantum Volume (depth=10) on IBM Power System AC922 with 6 GPUs (strong scaling). We used 6 GPUs per node for all cases, and for 1 node and 2 nodes we also stored in the memory of CPU.}
\label{fig:strong_34qubits}
\end{figure}

Next, we evaluated our cache blocking implementation on the parallel nodes of an IBM Power System AC922 with up to sixteen nodes. Fig. \ref{fig:strong_30qubits} and Fig. \ref{fig:strong_34qubits} are the measured simulation times of two Quantum Volume circuits. Fig. \ref{fig:strong_30qubits} shows a comparison of 30-qubit Quantum Volume calculated on two to sixteen nodes with only one GPU per node. Because the required calculation is relatively small, the simulation time is largely affected by data communications between nodes. When we add a node, the simulation time becomes worse in both implementations. But the cache blocking implementation has a smaller performance degradation. Fig. \ref{fig:strong_34qubits} compares the times of a 34-qubit Quantum Volume when chunks are stored on both the GPU and CPU for the one- and two-node cases. Because this circuit has enough calculations, the effect of cache blocking is relatively larger, and the simulation time with cache blocking is significantly shorter than that of the baseline. 

\begin{figure}[tbp]
  \centering
  \includegraphics[width=\linewidth]{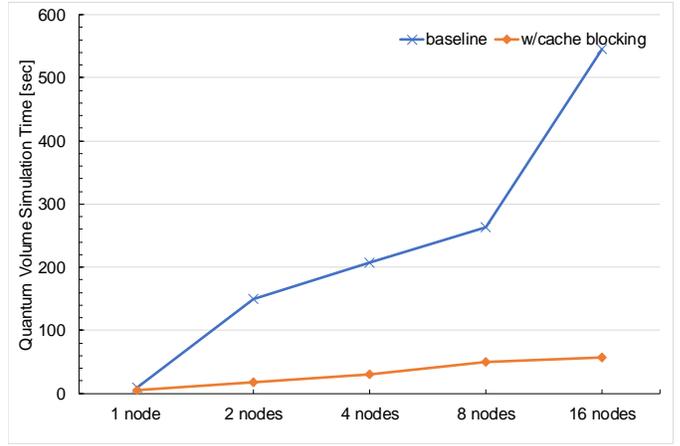}
\caption{Simulation time of Quantum Volume (depth=10) on IBM Power System AC922 with 6 GPUs and each node has $2^{30}$ amplitudes (weak scaling), i.e., 30 qubits for 1 node, 31 qubits for 2 nodes ...}
\label{fig:weak_scale_30}
\end{figure}

\begin{figure}[tbp]
  \centering
  \includegraphics[width=\linewidth]{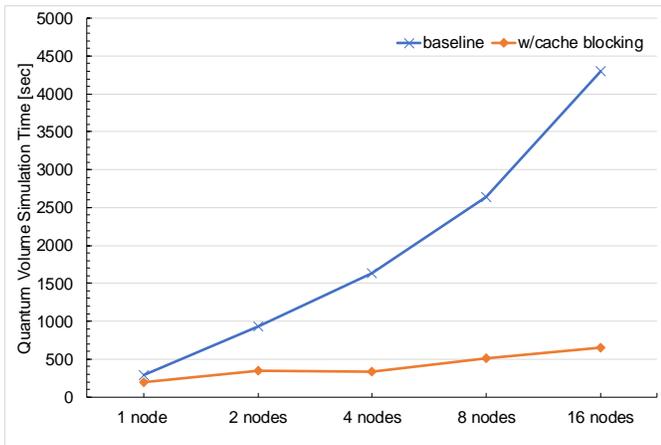}
\caption{Simulation time of Quantum Volume (depth=10) on IBM Power System AC922 with 6 GPUs and each node has $2^{33}$ amplitudes (weak scaling), i.e., 33 qubits for 1 node, 34 qubits for 2 nodes ...}
\label{fig:weak_scale_33}
\end{figure}

We also tested weak scalability by fixing the per node data size. Fig. \ref{fig:weak_scale_30} and Fig. \ref{fig:weak_scale_33} show weak scaling on the parallel nodes of an IBM Power System AC922 for up to sixteen nodes. In Fig. \ref{fig:weak_scale_33}, the per node array size is fixed to $2^{30}$ and six GPUs are used. This size is relatively small, meaning that the performance is significantly affected by the data transfer time between nodes; the baseline implementation shows a large scalability degradation. On the other hand, the cache blocking implementation shows better scalability. In Fig. \ref{fig:weak_scale_33}, there are $2^{34}$ amplitudes per node, and this requires both GPUs and CPU to store. In this case as well, the cache blocking implementation shows very good scalability.

\section{Summary}
\label{sec:summary}
We proposed optimization techniques for parallel quantum computing simulations. We decreased data exchanges by moving all the gates of the circuits to lower qubits by inserting noiseless swap gates. This resembles cache blocking on classical computers. Also, by combining chunk-based parallelization with cache blocking, we found that we can parallelize simulators with small changes because we can reuse serial simulation codes except for swap gates between chunks. This contributes to the productivity of the simulation software. 

We implemented a parallel state vector simulator in the open source quantum computing simulation framework, Qiskit Aer. We used the Thrust library instead of writing CUDA kernel codes to accelerate simulation on the GPUs. Since Qiskit Aer is written in C++ and based on STL, Thrust extends Qiskit Aer to support GPU naturally, and it contributes to productivity and portability. We also parallelized the state vector simulator by using MPI for hybrid parallel computers.

By applying parallelization techniques and the cache blocking technique described in this paper, we improved the scalability and simulation performance on a hybrid parallel computer. We improved both strong and weak scaling compared with previous implementations without cache blocking. The main contribution to the improvement is the decrease in data exchanges between chunks especially those on different nodes on a distributed parallel computer. Moreover, by storing chunks in the both the memory of the GPUs and the memory of the CPU, we can use the memory of the GPUs as a cache memory to accelerate simulations using the chunks stored in the CPU. 

We are planning to make this implementation public in the Qiskit Aer package. We are also planning to parallelize the rest of the simulation methods in Qiskit Aer, and we will evaluate them on hybrid parallel computers. To increase the number of qubits that can be be simulated, we can add slower storage such as NVMe to the node or network storage connected to the parallel cluster. The cache blocking technique can also improve performance by fetching chunks from these slower storages to the faster local storage and performing operations there and using blocking to decrease data movements from the slower storage.

\bibliographystyle{IEEEtran}
\bibliography{IEEEabrv,mybib}

\end{document}